**Millimeter-scale freestanding superconducting infinite-layer nickelate membranes**

Yonghun Lee[1,2,§,*], Xin Wei[1,3,§,*], Yijun Yu[1,2], Lopa Bhatt[4], Kyuho Lee[1,3,6], Berit H. Goodge[4,5,7], Shannon P. Harvey[1,2], Bai Yang Wang[1,3], David A. Muller[4,5], Lena F. Kourkoutis[4,5], Wei-Sheng Lee[1], Srinivas Raghu[1,3], and Harold Y. Hwang[1,2,*]

[1]Stanford Institute for Materials and Energy Sciences, SLAC National Accelerator Laboratory, Menlo Park, CA 94025, USA

[2]Department of Applied Physics, Stanford University, Stanford, CA 94305, USA

[3]Department of Physics, Stanford University, Stanford, CA 94305, USA

[4]School of Applied and Engineering Physics, Cornell University, Ithaca, New York 14850, USA

[5]Kavli Institute at Cornell for Nanoscale Technology, Cornell University, Ithaca, New York 14850, USA

[6]Present address: Department of Physics, Massachusetts Institute of Technology, Cambridge, MA 02139, USA

[7]Present address: Max Planck Institute for Chemical Physics of Solids, Dresden, Germany

*Email: yonghunl@stanford.edu (Y.L.); xinwei98@stanford.edu (X.W.); hyhwang@stanford.edu (H.Y.H.)

[§]Y.L. and X.W. contributed equally to this work.

**Abstract**

Progress in the study of infinite-layer nickelates has always been highly linked to materials advances. In particular, the recent development of superconductivity via hole-doping was predicated on the controlled synthesis of Ni in a very high oxidation state, and subsequent topotactic reduction to a very low oxidation state, currently limited to epitaxial thin films. Here we demonstrate a process to combine these steps with a heterostructure which includes an epitaxial soluble buffer layer, enabling the release of freestanding membranes of $(Nd,Sr)NiO_2$ encapsulated in $SrTiO_3$, which serves as a protective layer. The membranes have comparable structural and electronic properties to that of optimized thin films, and range in lateral dimensions from millimeters to ~100 micron fragments, depending on the degree of strain released with respect to the initial substrate. The changes in the superconducting transition temperature associated with membrane release are quite similar to those reported for substrate and pressure variations, suggestive of a common underlying mechanism. These membranes structures should provide a versatile platform for a range of experimental studies and devices free from substrate constraints.

**Keywords**

Nickelates, superconductivity, thin films, topochemical reduction, complex-oxide freestanding membranes

**Introduction**

In recent years, the application of soft chemistry techniques to oxide thin films and heterostructures has provided new routes to synthesize novel materials and establish unique sample geometries. The discovery of superconductivity in hole-doped infinite-layer nickelate thin films[1] utilized topotactic reduction to de-intercalate planes of oxygen from a parent perovskite phase. This process was demonstrated first in bulk powders[2,3] and subsequently in epitaxial thin films[4,5], where the substrate was essential to stabilizing the infinite-layer structure with uniform crystalline orientation. Another important example has been the creation of freestanding single crystal oxide membranes, either by dissolving the substrate itself[6,7] or a sacrificial layer underlying the membrane in a heterostructure[8,9].

This work is an attempt to combine these two soft chemistry techniques to create superconducting freestanding nickelate membranes. There are myriad motivations, fundamentally centered around a greater degree of materials control, to further study and utilize this new family of unconventional superconductors. One longstanding issue is that superconductivity in the infinite-layer nickelate has only been observed in thin film, not bulk form[10,11]. While exciting progress has been made in the synthesis of bulk single crystals[12], current limitations in the degree of accessible hole doping leave open the question whether epitaxial strain, or some other factor, explains the underlying distinction. Substrate variations have indicated that the superconducting transition temperature $T_c$ can be varied by strain[13], but the sensitivity to the formation of extended defects under varying strain conditions makes it difficult to unravel the impact of strain versus disorder[14]. Nevertheless, high pressure studies of thin films indicate that the scale of $T_c$ could exceed 30 K[15]. Freestanding membranes of this material would be valuable to address this issue, with potential to vary the strain state within the same sample[16].

In this study, we report the routine fabrication of freestanding infinite-layer nickelate membranes with 2 mm × 2 mm lateral dimensions, utilizing a water-soluble pseudo-perovskite $(Sr,Ca)_3Al_2O_6$ sacrificial layer[8]. The crystallinity and transport properties of these membranes are on par with thin film counterparts, establishing infinite-layer nickelate membranes as a viable platform to probe and manipulate superconducting and other physical properties.

**Synthesis strategies for freestanding infinite-layer nickelate membranes**

In order to obtain large-area freestanding oxide membranes utilizing the epitaxial pseudo-perovskite as a water-soluble sacrificial layer, the overall lattice matching throughout the heterostructure of the substrate, the sacrificial layer, and the target film is crucial for the minimization of strain gradients[16,17]. Without reasonable control of the epitaxial lattice mismatch, the strain energy released during the etching of the sacrificial layer is prone to generating cracks and wrinkles in the membrane. The synthesis of the infinite-layer nickelate membrane poses an additional challenge in this context because of the substantial expansion of the in-plane lattice constant of the nickelate during topochemical reduction, as well as the metastable nature of the infinite-layer nickelate phase.

One approach would be to first lift off the perovskite nickelate membrane and then reduce it to obtain an infinite-layer nickelate membrane, given that the infinite-layer nickelate phase is susceptible to re-oxidation and degradation from ambient oxygen and water[18,19]. In addition, the direct synthesis of perovskite nickelate membranes enables focus on lattice matching the perovskite alone, which has been shown to considerably impact the overall crystallinity[14]. However, we found that despite the ability to synthesize high-quality doped $Nd_{0.85}Sr_{0.15}NiO_3$ membranes, the subsequent topochemical reduction resulted in insulating transport properties and very low crystallinity (Figure S1, Supporting Information). While this result may not be definitive, we therefore directed our efforts to first create the infinite-layer phase before release in membrane form.

Several considerations arise for this approach. First, the criterion of minimizing epitaxial mismatch needs to be centered on the infinite-layer $Nd_{1-x}Sr_xNiO_2$, which is well-matched by the $SrTiO_3$ (STO) substrate and $Sr_2Ca_1Al_2O_6$ (SCAO) sacrificial layer[20] so that the membrane lift off proceeds with a modest release of ~0.5% compressive strain[3]. However, the $Nd_{1-x}Sr_xNiO_3$ initially grown has a large epitaxial mismatch of ~2.6% tensile strain[21]. As a point of comparison, we also studied membranes synthesized on $NdGaO_3$ (NGO) substrates using a $Sr_{0.5}Ca_{2.5}Al_2O_6$ sacrificial layer[20], which better matches the perovskite $Nd_{1-x}Sr_xNiO_3$ but results in higher (~1.6%) compressive strain[3] with respect to $Nd_{1-x}Sr_xNiO_2$. Another consideration is to minimize the risk of degradation of the infinite-layer nickelate phase during membrane release, where the film is exposed to water, oxygen, and mild heating. To address this, STO capping layers (8-20 unit cells (u.c.)) were epitaxially grown on both sides of the nickelate layer, to encapsulate and protect the nickelate from the environment[22].

**Growth, topochemical reduction, and membrane release**

We start from the epitaxial growth of 10 u.c. of STO / 22 u.c. of $Nd_{0.80}Sr_{0.20}NiO_3$ / 10 u.c. of STO / 10 u.c. of SCAO on a $TiO_2$-terminated STO substrate, as depicted in Figure 1a. In Figure 2a, the x-ray diffraction (XRD) $\theta$-$2\theta$ symmetric scan highlights (00$l$) reflections of $Nd_{0.80}Sr_{0.20}NiO_3$, surrounded by substantial Laue thickness oscillations. The close match of the XRD scan with a kinematic simulation[23] (see Experimental Section) confirms that the SCAO and STO layers are single-crystalline and grown to the desired thicknesses. Reciprocal space mapping (RSM) near the STO (103) reflection in Figure 2e confirms that the entire STO / $Nd_{0.80}Sr_{0.20}NiO_3$ / STO / SCAO heterostructure is coherently strained to the STO substrate.

A prerequisite for the realization of superconductivity in the infinite-layer nickelate phase is detailed control of the cation stoichiometry in its precursor perovskite phase[18]. Owing to the large epitaxial mismatch and the low thermodynamic stability associated with the high oxidation state of Ni in the perovskite, the epitaxy of $Nd_{1-x}Sr_xNiO_3$ on the STO substrate is highly susceptible to the formation of extended defects, most prominently identified as Ruddlesden-Popper-type vertical stacking faults[18,24,25]. On that account, two metrics obtained from the XRD $\theta$-$2\theta$ symmetric scan in Figure 2a – the strong (001) peak intensity and the extracted $c$-axis lattice constant of $3.710 \pm 0.005$ Å from the simulation (reported $c$-axis lattice constants are generally larger[1,18,25], attributed to the inclusion of vertical stacking faults) – point to the low density of extended defects, and hence the well-controlled cation stoichiometry, in $Nd_{0.80}Sr_{0.20}NiO_3$ grown on 10 u.c. of STO / 10 u.c. of SCAO / STO substrate (from here on referred to as the STO / SCAO template). Another challenge with the epitaxy of $Nd_{1-x}Sr_xNiO_3$ specific to this study is the sensitivity we observed with respect to the underlying STO / SCAO template. An atomically smooth step-and-terrace surface with no visible surface reconstructions is critical, as shown in the atomic force microscopy (AFM) image in Figure 1f and the reflection high-energy electron diffraction (RHEED) pattern in Figure 1e (see also Figure S2, Supporting Information). $Nd_{0.80}Sr_{0.20}NiO_3$ grown in nominally the same condition on a sub-optimal (typically partial layer completion) STO / SCAO template suffers from the inclusion of a much larger density of extended defects (Figure S3 and S4, Supporting Information).

Having established the growth of highly crystalline heterostructures, we employed calcium hydride for topochemical reduction (Figure 1b) as used in many prior studies[1,26]. The XRD $\theta$-$2\theta$ symmetric scan of the reduced heterostructure shown in Figure 2b demonstrates the effectiveness of this technique for films grown on the STO / SCAO template. The rightward shift of the Nd$_{0.80}$Sr$_{0.20}$NiO$_3$ (00$l$) reflections signifies a collapse of the $c$-axis lattice by the selective removal of apical oxygens with minimal film decomposition, which leads to a coherently $c$-axis oriented infinite-layer phase. The XRD simulation extracts a $c$-axis lattice constant of 3.329 ± 0.005 Å and a nickelate thickness of ~21 u.c. This perovskite-to-infinite-layer conversion ratio close to unity suggests a nearly complete transition to the infinite-layer phase with minimal decomposition, especially considering the interface layers. The RSM in Figure 2f further shows that heterostructure post-reduction remain fully strained to the STO substrate. The AFM image of the reduced heterostructure in Figure 1g shows a step-and-terrace morphology, consistent with the layer-by-layer growth mode of the precursor perovskite and the preservation of an atomically flat surface after the reduction process.

The robustness of the sacrificial layer under the strongly reducing environment is noteworthy, and of course central to this approach to synthesizing superconducting nickelate membranes. The XRD simulation for the reduced heterostructure in Figure 2b is performed with the same lattice and thickness parameters of the SCAO as prior to the reduction. The close correspondence with the XRD scan thus supports our schematic assumption in Figure 1b, in which the region of topochemical reduction is confined to the nickelate layer. Importantly, the SCAO retains its solubility in water (Figure S5, Supporting Information).

By dissolving the SCAO sacrificial layer in water with polymer supports in place (Figure 1c, see Experimental Section), we obtain a millimeter-scale, uniformly continuous Nd$_{0.80}$Sr$_{0.20}$NiO$_2$ freestanding membrane, as shown in the optical microscope image in Figure 1h. The size and uniformity highlight the importance of the lattice-matched heterostructure for coherent membrane release over large areas. By contrast, Nd$_{0.85}$Sr$_{0.15}$NiO$_2$ membranes released from the NGO substrate (with greater strain mismatch of ~1.6%) exhibit cracks and wrinkles, which limit the size of continuous membranes to below ~200 μm (Figure S6, Supporting Information).

To examine the structural integrity of the infinite-layer nickelate phase detached from the substrate, we carry out XRD $\theta$-$2\theta$ symmetric scans on a freestanding STO / Nd$_{0.80}$Sr$_{0.20}$NiO$_2$ / STO heterostructure membrane transferred onto a Si wafer covered with a 300-nm-thick SiO$_2$ surface layer (Figure 1d). In Figure 2c, two membrane XRD peaks are visible, one from the (002) reflection of STO and another from Nd$_{0.80}$Sr$_{0.20}$NiO$_2$, at a similar peak position as that of the parent thin film heterostructure before release. A side-by-side comparison with the XRD scan of the pristine film reveals that a series of small shoulder peaks connecting STO (002) and Nd$_{0.80}$Sr$_{0.20}$NiO$_2$ (002) reflections track the Laue oscillations generated by the single-crystalline phases of the heterostructure. Notably, the peak position of Nd$_{0.80}$Sr$_{0.20}$NiO$_2$ (002) shifts rightward, indicating a decrease in the *c*-axis lattice constant due to the relaxation from its initially epitaxially compressed state on the STO substrate. The change in the *c*-axis lattice constant is negligible within measurement uncertainty – as expected, since not only is the bulk in-plane lattice constant of Nd$_{1-x}$Sr$_x$NiO$_2$ close to that of STO (ref. [3]), but also the freestanding nickelate is still partially strained by the surrounding STO layers.

On the other hand, the release of a 8 u.c. STO / 14 u.c. Nd$_{0.85}$Sr$_{0.15}$NiO$_2$ / 8 u.c. STO membrane from the NGO substrate exhibits a more significant *c*-axis relaxation of 0.75 ± 0.30%, from 3.323 ± 0.005 Å to 3.298 ± 0.01 Å (Figure 2d), due to a larger epitaxial compressive strain of ~1.6% (ref. [3]) up to the post-reduction stage, as shown in the RSM scans in Figures 2g-h. (See

Figure S7, Supporting Information, for the XRD characterization of $Nd_{0.85}Sr_{0.15}NiO_3$ and $Nd_{0.85}Sr_{0.15}NiO_2$ thin films grown on STO / $Sr_{0.5}Ca_{2.5}Al_2O_6$ / NGO substrate.)

**Scanning transmission electron microscopy of the infinite-layer nickelate membranes**

Figures 3a-d show high-angle annular dark-field (HAADF) scanning transmission electron microscopy (STEM) images of a STO / $Nd_{0.8}Sr_{0.2}NiO_2$ / STO membrane transferred onto a silicon nitride ($SiN_x$) TEM window. Comparing to previous studies of $Nd_{1-x}Sr_xNiO_2$ thin films grown on STO substrates[18,24], the cross-sectional STEM image in Figure 3a shows a highly crystalline structure with a surprisingly low density of extended defects, consistent with what is inferred from the XRD characterization presented above. The ADF image and elemental electron energy loss spectroscopy (EELS) mapping in Figure 3d show sharp interfaces with limited signs of interdiffusion beyond the single-unit-cell reconstruction[27,28] across the two interfaces of STO and $Nd_{0.80}Sr_{0.20}NiO_2$. Crucially, the overall homogeneity of the STEM image, along with the macroscopic uniformity observed in the optical microscope image of the membrane (Figure 1h), indicates the robustness of the heterostructure, particularly the nickelate layer, to the membrane fabrication process. The membrane geometry itself provides an attractive and direct sample for plan-view imaging, as shown in the HAADF STEM image in Figure 3b. Together with the strain mapping shown in Figure 3c highlighting the Ruddlesden-Popper faults (see Experimental Section), the overall structure of the extended defects can be visualized and shows that the domains of faults are sparse and isolated from each other.

**Superconductivity in infinite-layer nickelate membranes**

The temperature-dependent resistivity $\rho(T)$ of the reduced $Nd_{0.80}Sr_{0.20}NiO_2$ thin film heterostructure is shown in Figure 4a, exhibiting transport signatures characteristic of optimally doped samples with high crystallinity[1,14,26]: reasonable metallicity with a residual resistivity $\rho(20$ K$)$ of 0.10 m$\Omega$ cm, a residual resistivity ratio $\rho(290$ K$)/\rho(20$ K$)$ of 4.2, and a superconducting transition at $T_{c,onset}$ = 17.3 K (defined as temperature at which the second derivative of $\rho(T)$ becomes negative) that hits the measurement noise floor at $T_{c,0}$ = 12.0 K. An important finding of this study is the very close similarity of $\rho(T)$ of the released $Nd_{0.80}Sr_{0.20}NiO_2$ membrane (Figure 4a) with only a slight decrease in $T_{c,onset}$ = 17.0 K and $T_{c,0}$ = 11.7 K.

Overall, little to no sign of macroscopic degradation is found in the electrical properties of the membranes, complementing the structural findings via XRD and STEM. Furthermore, the overall decrease of $T_c$ by ~0.3 K after lift off is not concomitant with a broadening of the transition, suggesting that the changes are dominated by a change in the strain state. We also performed magnetotransport measurements on the $Nd_{0.80}Sr_{0.20}NiO_2$ membrane. Figure 4b shows a series of $\rho(T)$ curves taken with increasing perpendicular field strengths up to 14 T. While the normal state $\rho(T)$ above $T_c$ exhibits a negligible field dependence, upon entering the superconducting state the magnetoresistive transitions are suppressed and broadened as found in thin films.

The large area of the $Nd_{0.80}Sr_{0.20}NiO_2$ membranes released from the STO substrate, such as the one in the inset of Figure 4c (which is actually a membrane cut in half to fit the measurement), enables magnetic measurements with a two-coil mutual-inductance probe optimized for the thin film geometry[29] – in particular the diamagnetic response in the superconducting state. Figure 4c shows the temperature dependence of the real ($Re(V_p)$) and imaginary ($Im(V_p)$) component of the AC voltage signal $V_p$, measured from the pickup coil on the other side of the membrane / Si wafer from the coaxial drive coil (see Experimental Section). $Im(V_p)$ exhibits a dissipation peak, the width of which is a measure of the superconducting transition on the scale of the dimensions (~mm) probed by the coil[29]. The relatively sharp peak with a width of ~2 K is similar to that observed in optimized films, consistent with what is inferred from the resistive transition (Figure 4a). $Re(V_p)$ represents the diamagnetic screening of the magnetic field from the drive coil, which increases with decreasing temperature with growth of the superfluid density.

**Strain response of $T_c$ in the infinite-layer nickelate**

A more pronounced decrease of $T_c$ is found in the membrane released from the NGO substrate. $\rho(T)$ of the precursor $Nd_{0.85}Sr_{0.15}NiO_2$ film in Figure 4d shows that $T_{c,onset}$ ($T_{c,0}$) reaches as high as 25.7 K (20.7 K). This is rather noteworthy in and of itself, in being the highest reported $T_c$ for the (Nd,Sr)NiO$_2$ system to the best of our knowledge. This is also consistent with the $T_c$ increase attributed to compressive strain for $Pr_{0.8}Sr_{0.2}NiO_2$ induced by substrate variations[13]. $\rho(T)$ of the released membrane (measurement geometry shown in the inset of Figure 4d), however, shows that $T_{c,onset}$ ($T_{c,0}$) reduces to 18.0 K (11.9 K) – a ~40% drop, significantly larger than the case of the membrane released from the STO substrate.

These observations motivate us to compare our current results with previous reports on (Pr or Nd)$_{1-x}$Sr$_x$NiO$_2$ thin films in Figure 5, focusing on the possible correlation of $T_c$ and strain, previously mediated by varying the substrate-induced epitaxial strain[13,14] and by pressure applied to thin film samples[15]. (Note that for the high-pressure results[15], the nickelate in-plane lattice constants are assumed to follow that of the STO substrate, compressing under isotropic pressure via the bulk STO response[30].) It should again be emphasized that the degree of disorder is likely varying across the samples of this comparison, as discussed above, as well as the specific hole-doping values (here chosen in the vicinity of optimal doping). Despite these caveats, a rather uniform rate of increase in $T_c$ is found with decreasing in-plane lattice constant, indicating that $T_c$ is increasing with larger electronic bandwidth. The microscopic origin of this trend, and the ultimate limit on superconductivity that is implied, is an important question for further investigation.

**Conclusions**

In summary, we have demonstrated the synthesis of highly crystalline and superconducting infinite-layer nickelate freestanding membranes. For the closely lattice-matched heterostructures grown on STO, millimeter-scale continuous membranes could be reproducibly grown. For the heterostructures grown under significant compressive strain on NGO, crack formation during membrane release limited typical membrane fragments to the ~0.1 millimeter scale. In both cases, the electrical properties were comparable to those of optimal thin film samples.

Opportunities abound for future experiments using infinite-layer nickelate membranes. To further investigate their strain response, a next step is to study and establish their mechanical properties, such as fracture strain limits[31]. If a wide regime of elastic strain is found as in some other oxide membranes[16,32], variations in the superconducting and normal state properties can be probed under tensile strain. Establishing whether $T_c$ is varying due to an overall change in bandwidth, or via specific channels such as Nd 5*d* - Ni 3*d* hybridization or O 2*p* - Ni *3d* hybridization, would be particularly insightful[33–35]. Furthermore, the membrane geometry will enable a range of experiments which are currently technically limited by the presence of a substrate.

**Experimental Section**

*Thin-Film Growth*: 8-20 u.c. SrTiO$_3$ / 15-25 u.c. Nd$_{1-x}$Sr$_x$NiO$_3$ ($x$ = 0.15 and 0.20) / 8-20 u.c. SrTiO$_3$ / 10-15 u.c. Sr$_{3-x}$Ca$_x$Al$_2$O$_6$ ($x$ = 1.0 and 2.5) heterostructures were grown on 5 × 5 mm$^2$ (001)-oriented single-crystalline SrTiO$_3$ substrates and (110)-oriented NdGaO$_3$ substrates by pulsed laser deposition (KrF, $\lambda$ = 248 nm). SrTiO$_3$ substrates were *in situ* pre-annealed at 930 °C under oxygen partial pressure $P_{O_2}$ = 5 × 10$^{-6}$ Torr for 30 minutes to obtain a step-and-terrace surface, while NdGaO$_3$ substrates were pre-annealed at 1000 °C for 2 hours in air. The sacrificial layer Sr$_{3-x}$Ca$_x$Al$_2$O$_6$ was grown at 700 °C with $P_{O_2}$ = 5 × 10$^{-6}$ Torr, laser fluence $F$ = 3.12 J cm$^{-2}$ (spot size = 2.20 mm$^2$) and laser repetition frequency $f$ = 1 Hz. SrTiO$_3$ buffer layers were grown at 700 °C with $P_{O_2}$ = 5 × 10$^{-6}$ Torr, $F$ = 0.89 J cm$^{-2}$ (spot size = 3.20 mm$^2$), and $f$ = 1 Hz. Nd$_{1-x}$Sr$_x$NiO$_3$ thin films were grown at 580 °C with $P_{O_2}$ = 150 mTorr, $F$ = 2.37 J cm$^{-2}$ (spot size = 2.38 mm$^2$), and $f$ = 2 Hz. Finally, the SrTiO$_3$ capping layers were grown under the same conditions as Nd$_{1-x}$Sr$_x$NiO$_3$ given the limited stability of the hole-doped perovskite nickelate. A single-crystal SrTiO$_3$ target was used for the SrTiO$_3$ growth, while polycrystalline ceramic targets were used for Sr$_{3-x}$Ca$_x$Al$_2$O$_6$ and Nd$_{1-x}$Sr$_x$NiO$_3$ growths.

*Topochemical Reduction*: The as-grown films were loosely wrapped with aluminum foil and inserted into a Pyrex glass tube along with ~0.15 g of calcium hydride (97% purity, Sigma-Aldrich). The tubes were pumped with a rotary vane pump for 10 minutes to reach a pressure below 5 mTorr, and then sealed with a hydrogen torch. The sealed glass tubes were subsequently put in a tube furnace and ramped to 330-360 °C at a rate of 10 °C min$^{-1}$, annealed for 2-6 hours, and cooled down to room temperature at a rate of approximately 2 °C min$^{-1}$.

*Thin Film Lift off and Membrane Transfer*: The reduced heterostructure films were spin-coated with A8 950 polymethyl methacrylate (PMMA) at a speed of 3500 rpm for 60 seconds with a ramping rate of 1000 rpm s$^{-1}$ (thickness ~1 μm) and then cured at 135 °C for 7.5 minutes. Sample edges were exposed by scratching with a tungsten pen where water could enter and start etching the sacrificial layer. Sheets of plasma-treated polydimethylsiloxane (PDMS) (~2 × 2 mm$^2$) were then applied as a mechanical support. The samples were then put in deionized water until the sacrificial layers had been fully dissolved. The released membranes were scooped out of the water with a 3 mm-radius metal ring and transferred onto a target substrate, such as a Si/SiO$_2$ wafer or a SiN$_x$ membrane with circular holes 2 μm in diameter (Norcada Inc.). The samples were then dried at 60 °C for 5 minutes and then subsequently 110 °C for 10 minutes. After heating, the samples were dipped into acetone at room temperature to dissolve the PMMA (~10 minutes), until the PDMS had naturally been released from the samples. Finally, the samples were dipped in isopropanol for 5 minutes to remove all other residues.

*Thin Film and Membrane Characterization*: X-ray diffraction (XRD) $\theta$-$2\theta$ symmetric scans and reciprocal space mapping (RSM) were measured using monochromatic Cu $K_{\alpha 1}$ radiation sources ($\lambda$ = 1.5406 Å). Temperature- and field-dependent resistivity $\rho(T, H)$ measurement of the thin film samples were conducted in a van der Pauw geometry with aluminum electrical wire-bonded contacts. For the millimeter-sized freestanding membrane samples, thin and narrow sheets of indium were pressed on the membrane in a four-point probe geometry, making a robust Ohmic contact at low temperature while penetrating through the thick STO capping layer. For the small flake samples, the STO capping layers were removed by argon ion-milling before making electrical contacts with Ti 5 nm / Au 40 nm electrodes. The mutual inductance probe setup used in ref. [29] was employed in this study. An infinite-layer nickelate membrane was transferred onto a Si wafer and then cut in half, which was placed in between two coils – a drive coil with 0.25 mm diameter and 50 turns and a pickup coil with 0.5 mm diameter and 400 turns.

*Scanning Transmission Electron Microscopy (STEM) and Analysis*: Electron transparent cross-sectional STEM samples were prepared with the standard focused ion beam (FIB) lift-out procedure using a Thermo Fisher Scientific Helios G4 UX focused ion beam. Annular dark field (ADF)-STEM was performed on an aberration-corrected FEI Titan Themis and Thermo Fisher Scientific Spectra 300 X-CFEG at an accelerating voltage of 300 kV with a convergence angle of 30 mrad and inner (outer) collection angles of 68 (340) mrad and 66 (200) mrad, respectively. Elemental maps were recorded on the same Titan Themis with a Gatan 965 GIF Quantum ER spectrometer and a Gatan K2 Summit direct electron detector operated in counting mode. Ruddlesden-popper faults in the plan-view ADF-STEM images of the film were visualized by extracting modulations in the (110) and (–110) pseudo-cubic lattice fringes using the phase lock-in method described in refs. [36,37]. In particular, the Ruddlesden-popper fault boundaries appear as an apparent strong local compressive strain in the pseudo-cubic lattice fringes. The strain map in Figure 3c is generated by taking the positive values of the sum of (110) and (–110) compressive strain map.

*Note*: During the preparation of this manuscript, we became aware of a report on $La_{0.80}Sr_{0.20}NiO_2$ freestanding membranes[38].

**Acknowledgements**

Y.L. and X.W. contributed equally to this work. The authors acknowledge D. Li and V. Harbola for their contribution to the initial stage of this work, W. J. Kim, E. K. Ko and A. Vailionis for assistance with RSM measurements, and E. K. Ko, J. Fowlie, K. J. Crust, and J. Wang for critical reading of the manuscript. This work was supported by the U. S. Department of Energy, Office of Basic Energy Sciences, Division of Materials Sciences and Engineering (Contract No. DE-AC02-76SF00515) and the Gordon and Betty Moore Foundation's Emergent Phenomena in Quantum Systems Initiative (Grant No. GBMF9072, synthesis equipment). Electron microscopy by L.B., B.H.G., D.A.M. and L.F.K. was supported by PARADIM through NSF DMR-2039380 with additional support by the Department of Defense Air Force Office of Scientific Research (Grant No. FA 9550-16-1-0305) and the Packard Foundation. This work made use of a Helios FIB supported by the NSF (Grant No. DMR-2039380) and the Cornell Center for Materials Research Shared Facilities, which are supported through the NSF MRSEC program (Grant No. DMR-1719875). The Thermo Fisher Spectra 300 X-CFEG was acquired with support from Platform for the Accelerated Realization, Analysis, and Discovery of Interface Materials (PARADIM), an NSF MIP (No. DMR-2039380) and Cornell University. The FEI Titan Themis was acquired through NSF-MRI-1429155, with additional support from Cornell University, the Weill Institute, and the Kavli Institute at Cornell.

**Conflict of Interest**

The authors declare no conflict of interest.

**Figures**

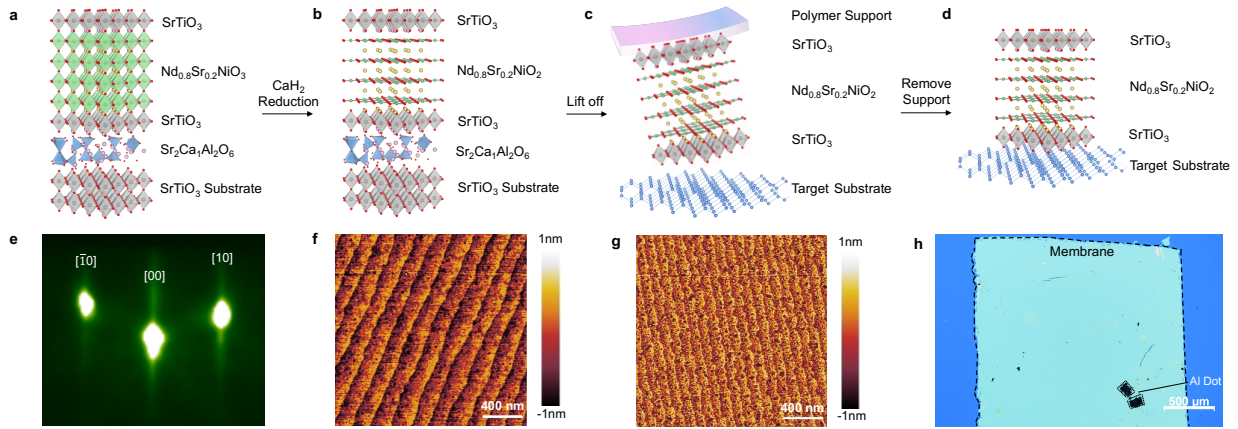

**Figure 1**. Schematic diagrams illustrating the synthesis of freestanding infinite-layer nickelate membranes and representative images of the sample at various steps. a) The STO / Nd$_{0.80}$Sr$_{0.20}$NiO$_3$ / STO / SCAO heterostructure grown on the STO substrate. b) Topochemical reduction – selectively reducing the nickelate layer. c) Thin film lift off. Polymer supports are used, comprising a sheet of PDMS attached on top of spin-coated PMMA to keep the STO / Nd$_{0.80}$Sr$_{0.20}$NiO$_2$ / STO heterostructure stable during the lift off process. d) The STO / Nd$_{0.80}$Sr$_{0.20}$NiO$_2$ / STO membrane transferred on a target substrate and the polymer supports removed. e) RHEED pattern of a STO buffer grown on SCAO recorded along the [100] azimuth. f) AFM image of the surface of STO / SCAO template. g) AFM image of the surface of the top STO capping layer for the reduced heterostructure schematically represented in (b). h) Optical microscope image of a STO / Nd$_{0.80}$Sr$_{0.20}$NiO$_2$ / STO membrane (~2 mm × 2 mm) transferred onto a Si wafer.

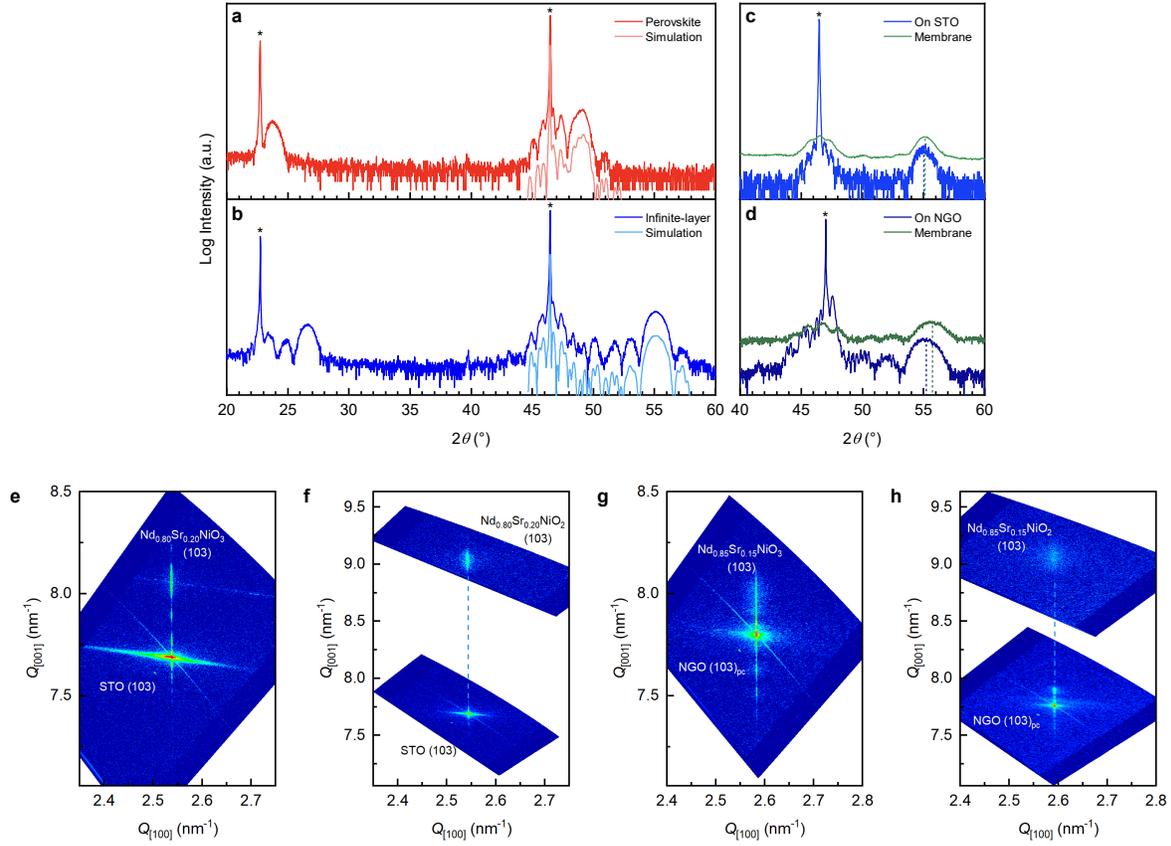

**Figure 2**. Structural characterization of nickelate thin film heterostructures and membranes. a) XRD $\theta$-$2\theta$ symmetric scan and corresponding simulation near the (002) Bragg peaks based on the kinematic model[23] of a 10 u.c., 3.905 Å of STO / 22 u.c., 3.710 Å of $Nd_{0.80}Sr_{0.20}NiO_3$ / 10 u.c., 3.905 Å of STO / 10 u.c., 15.65 Å of SCAO heterostructure grown on the STO substrate. * denotes substrate peaks. b) Sample in (a) after topochemical reduction. The simulation parameters are the same as those in (a), except that the parameters for $Nd_{0.80}Sr_{0.20}NiO_2$ is changed to 21 u.c., 3.329 Å. c) XRD $\theta$-$2\theta$ symmetric scans of a nickelate heterostructure grown on a STO substrate, and the released membrane transferred onto a Si wafer. Dashed lines mark the peak positions for the film and the membrane. d) XRD $\theta$-$2\theta$ symmetric scan of a 8 u.c. STO / 14 u.c. $Nd_{0.85}Sr_{0.15}NiO_2$ / 8 u.c. STO / 15 u.c. $Sr_{0.5}Ca_{2.5}Al_2O_6$ grown on the NGO substrate, and released in membrane form. e-f) RSM of $Nd_{0.80}Sr_{0.20}NiO_3$ and $Nd_{0.80}Sr_{0.20}NiO_2$, sandwiched by STO and grown on 10 u.c. SCAO / STO substrate. g-h) RSM $Nd_{0.85}Sr_{0.15}NiO_3$ and $Nd_{0.85}Sr_{0.15}NiO_2$, sandwiched by STO and grown on 15 u.c. $Sr_{0.5}Ca_{2.5}Al_2O_6$ / NGO substrate.

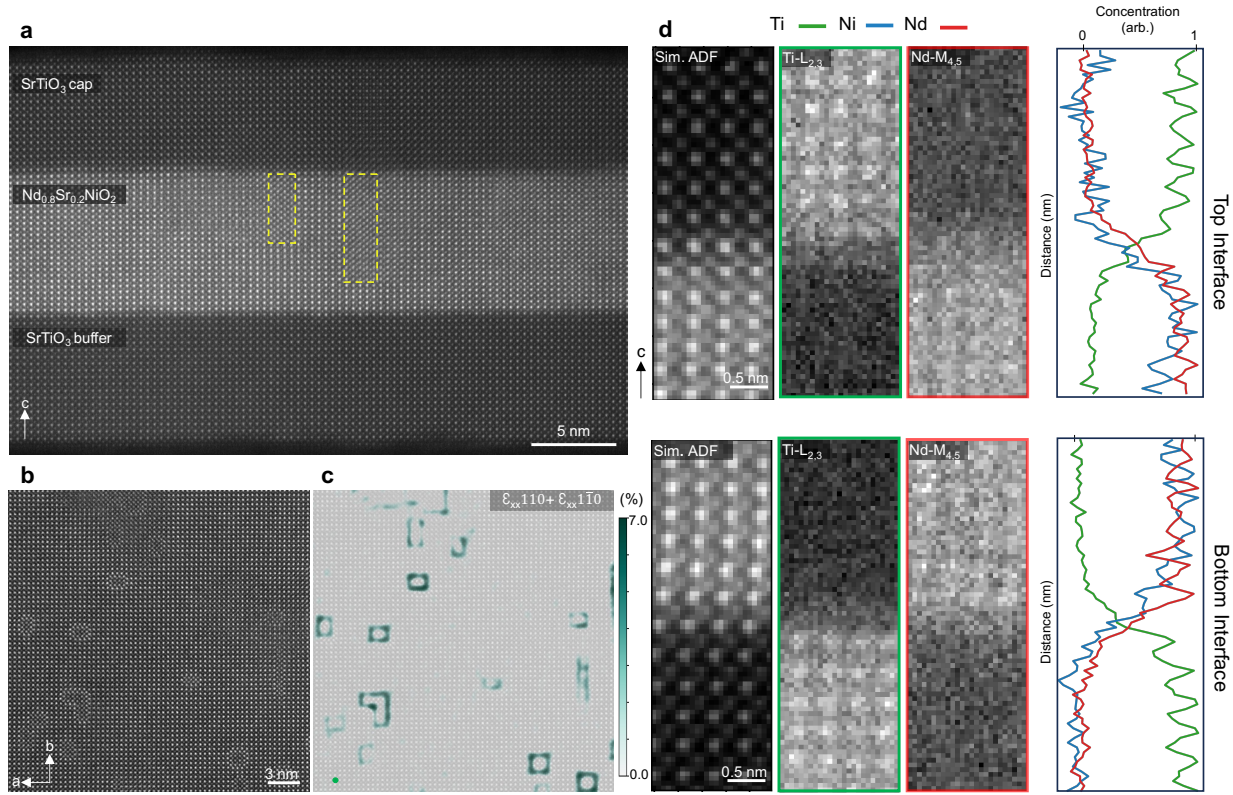

**Figure 3**. STEM images of a STO / $Nd_{0.8}Sr_{0.2}NiO_2$ / STO membrane transferred onto a $SiN_x$ TEM window. a) Cross-sectional HAADF STEM image. Yellow dashed lines outline regions of Ruddlesden-Popper faults. b) Plan-view HAADF STEM image. c) (±110) compressive strain map of the plane-view image in (b), highlighting the Ruddlesden-Popper faults. d) Cross-sectional ADF image acquired near the top and bottom interface of STO and $Nd_{0.8}Sr_{0.2}NiO_2$ (left-most column), along with corresponding elemental EELS maps of Ti-$L_{2,3}$ and Nd-$M_{4,5}$ (middle columns) with their intensity profiles in the right most plots (green, blue, and red, respectively).

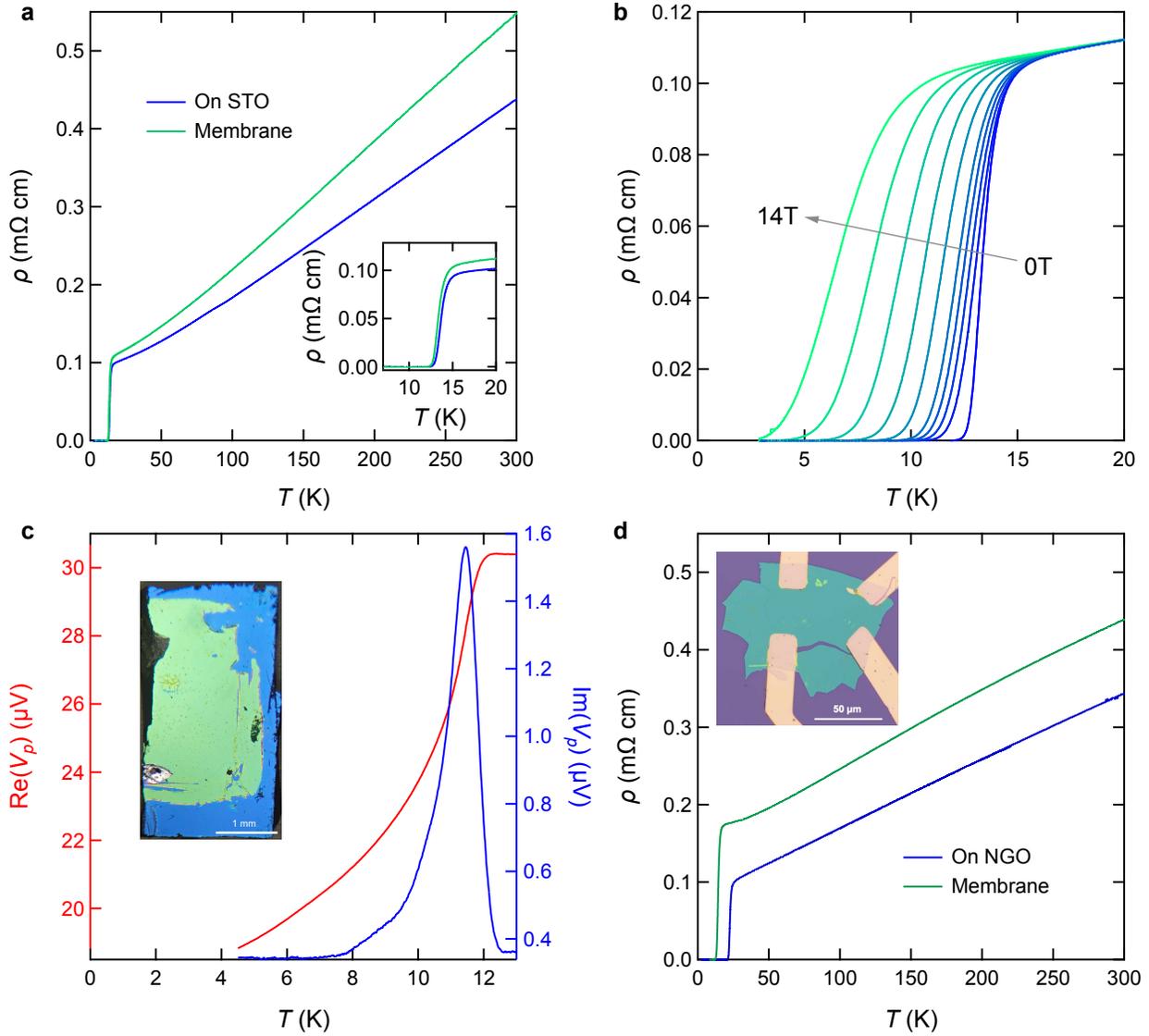

**Figure 4**. Electrical and magnetic characteristics of the infinite-layer nickelate heterostructures and membranes. a) $\rho(T)$ plots of a STO / Nd$_{0.80}$Sr$_{0.20}$NiO$_2$ / STO heterostructure grown on the STO substrate before and after transfer onto a Si wafer. The inset shows $\rho(T)$ near the superconducting transition. b) $\rho(T)$ measured under increasing perpendicular magnetic field. c) Temperature-dependent real and imaginary component of AC voltage measured in the pick-up coil of the mutual inductance measurement of a STO / Nd$_{0.80}$Sr$_{0.20}$NiO$_2$ / STO membrane transferred onto a Si wafer (shown in inset). d) $\rho(T)$ of a STO / Nd$_{0.80}$Sr$_{0.20}$NiO$_2$ / STO heterostructure grown on an NGO substrate, before and after transfer onto a Si wafer. The inset shows the optical microscope image of the membrane device geometry.

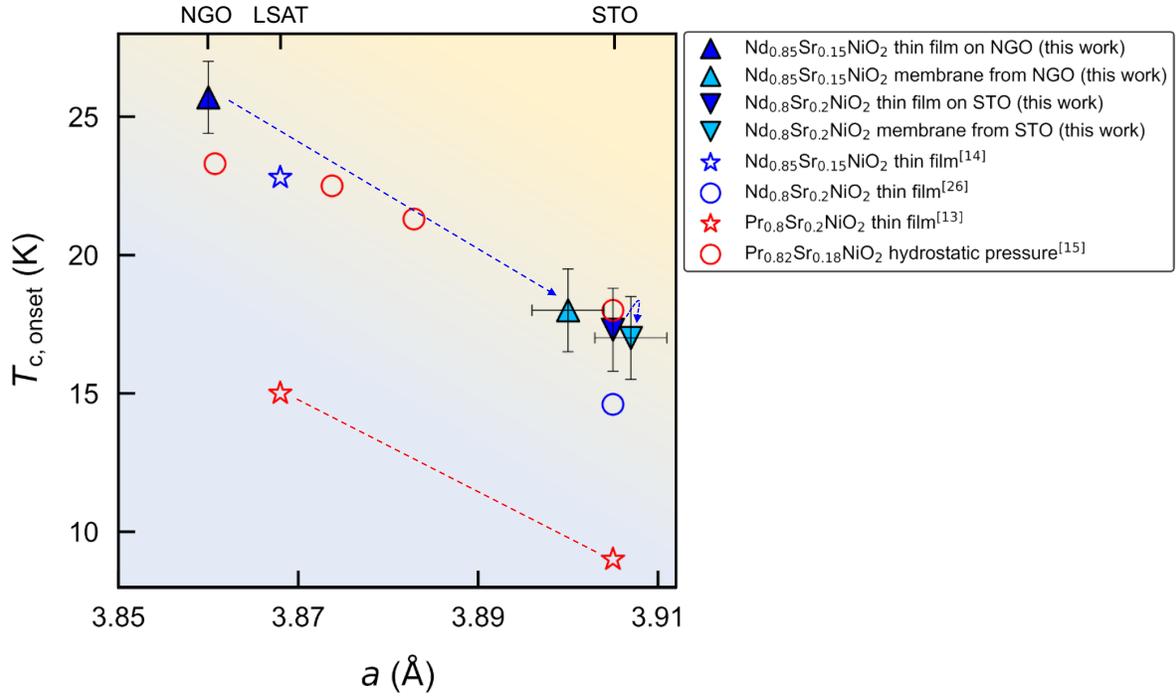

**Figure 5**. $T_{c,\,onset}$ versus $a$-axis lattice constant of (Nd or Pr)$_{1-x}$Sr$_x$NiO$_2$ in thin film and membrane form. In the case of high-pressure data[15], the lattice constants are assumed to follow that of the STO substrate and bulk Young's modulus[30]. Vertical error bars represent the spread of $T_c$ from sample-to-sample variations. Horizontal error bars represent the uncertainties in fitting the Nd$_{1-x}$Sr$_x$NiO$_2$ (200) reflection from grazing incident XRD of the membrane samples.

**Supporting Information for**

**Millimeter-scale freestanding superconducting infinite-layer nickelate membranes**

Yonghun Lee[1,2§*], Xin Wei[1,3§*], Yijun Yu[1,2], Lopa Bhatt[4], Kyuho Lee[1,3,6], Berit H. Goodge[4,5,7], Shannon P. Harvey[1,2], Bai Yang Wang[1,3], David A. Muller[4,5], Lena F. Kourkoutis[4,5], Wei-Sheng Lee[1], Srinivas Raghu[1,3], and Harold Y. Hwang[1,2*]

[1]*Stanford Institute for Materials and Energy Sciences, SLAC National Accelerator Laboratory, Menlo Park, CA 94025, USA*
[2]*Department of Applied Physics, Stanford University, Stanford, CA 94305, USA*
[3]*Department of Physics, Stanford University, Stanford, CA 94305, USA*
[4]*School of Applied and Engineering Physics, Cornell University, Ithaca, New York 14850, USA*
[5]*Kavli Institute at Cornell for Nanoscale Technology, Cornell University, Ithaca, New York 14850, USA*
[6]*Present address: Department of Physics, Massachusetts Institute of Technology, Cambridge, MA 02139, USA*
[7]*Present address: Max Planck Institute for Chemical Physics of Solids, Dresden, Germany*

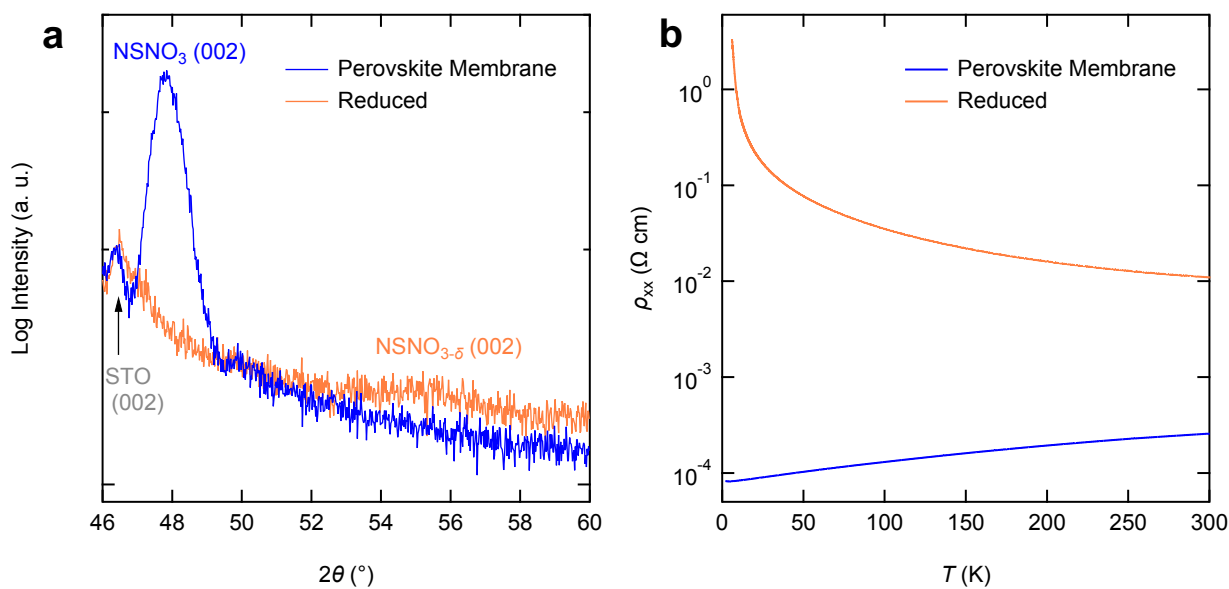

**Figure S1**. XRD and transport characterization of a perovskite nickelate $Nd_{0.85}Sr_{0.15}NiO_3$ ($NSNO_3$) membrane before and after topochemical reduction. a) XRD $\theta$-$2\theta$ symmetric scan of the transferred membrane in the perovskite phase (blue) and after reduction (orange). STO refers to the STO (002) reflection of $SrTiO_3$ capping layers. The clear $Nd_{0.85}Sr_{0.15}NiO_3$ (002) reflection of the perovskite membrane is contrasted by the negligible peak intensity after reduction. b) Temperature-dependent resistivity of the perovskite (blue) and reduced (orange) membrane.

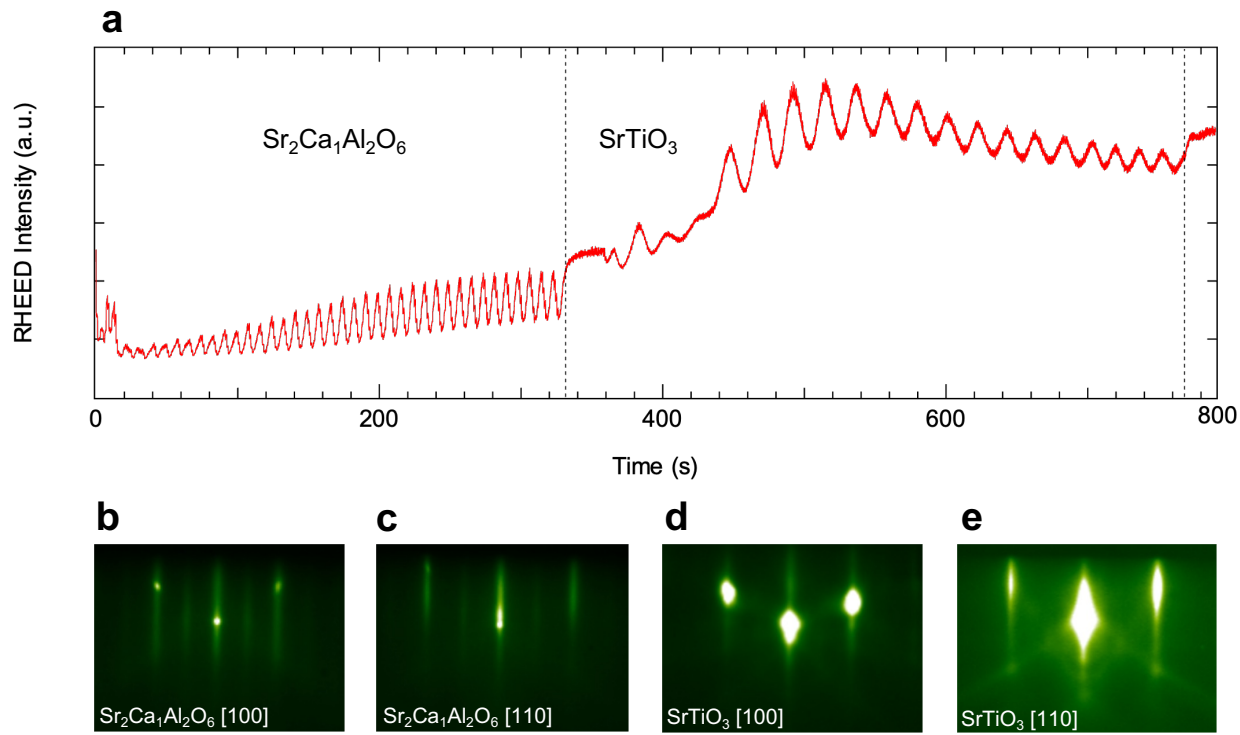

**Figure S2**. Additional RHEED characterization of the optimized STO / SCAO template. a) RHEED oscillations for the growth of 10 u.c. $Sr_2Ca_1Al_2O_6$ and 20 u.c. $SrTiO_3$. The periodicity of the oscillations for the $Sr_2Ca_1Al_2O_6$ growth is quadrupled, indicating the periodicity of the perovskite subunits. The dashed lines indicate the end of each stage of the growth. b-c) RHEED patterns of 10 u.c. $Sr_2Ca_1Al_2O_6$ grown on the STO substrate recorded along b) the [100] azimuth and c) the [110] azimuth. d-e) RHEED patterns of 10 u.c. STO grown on $Sr_2Ca_1Al_2O_6$ / STO substrate recorded along d) the [100] azimuth and e) the [110] azimuth.

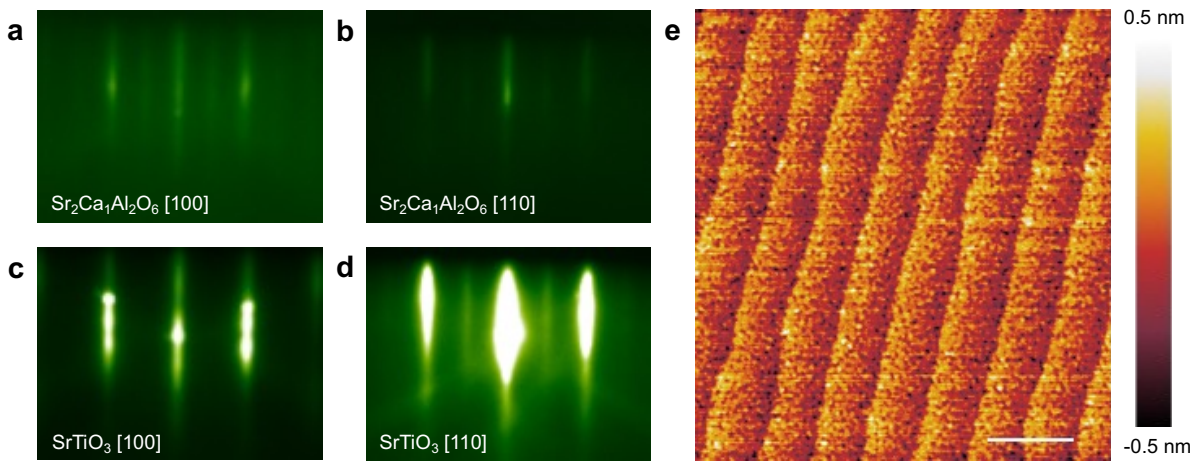

**Figure S3**. Surface characteristics of a sub-optimally grown STO/SCAO template. a-b) RHEED patterns of 10 u.c. $Sr_2Ca_1Al_2O_6$ grown on the STO substrate recorded along a) the [100] azimuth and b) the [110] azimuth. c-d) RHEED patterns recorded along c) the [100] azimuth and d) the [110] azimuth, along with e) the corresponding AFM scan of 10 u.c. STO grown on $Sr_2Ca_1Al_2O_6$ / STO substrate. Scale bar, 0.4 μm.

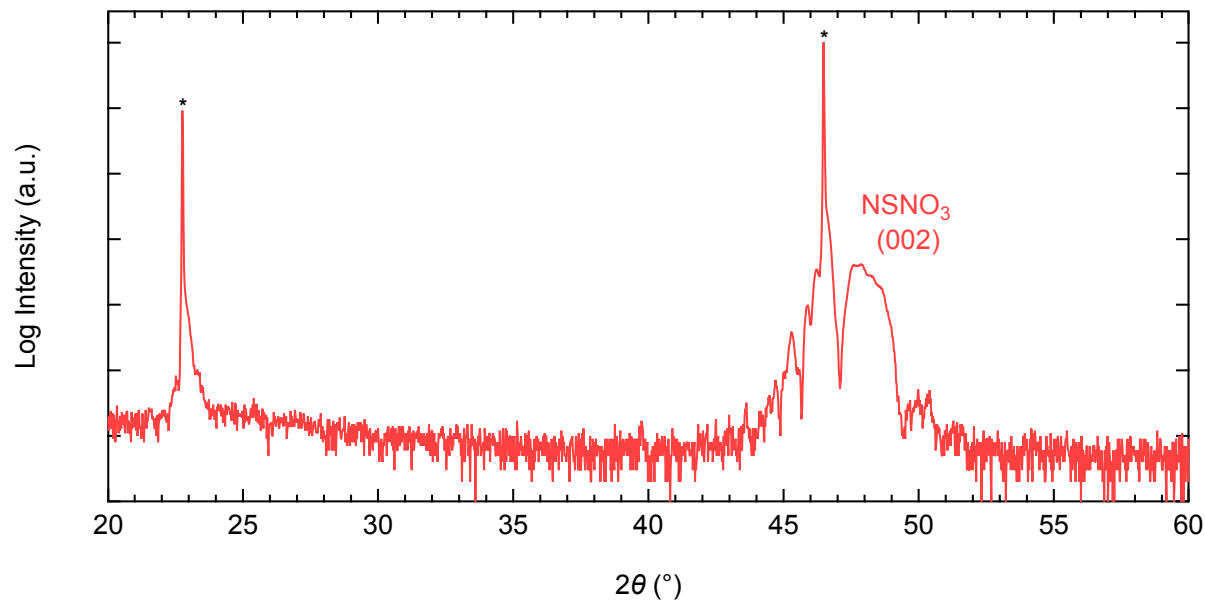

**Figure S4**. XRD $\theta$-$2\theta$ symmetric scan of $Nd_{0.80}Sr_{0.20}NiO_3$ grown on the sub-optimal STO / $Sr_2Ca_1Al_2O_6$ template (Figure S3) using the same nominal growth conditions as in the main text. The $Nd_{0.80}Sr_{0.20}NiO_3$ (001) is almost non-existent and the (002) position (~48.3°) is left-shifted compared to the optimized sample in the main text (~49.0°, Figure 2a). * denotes substrate peaks.

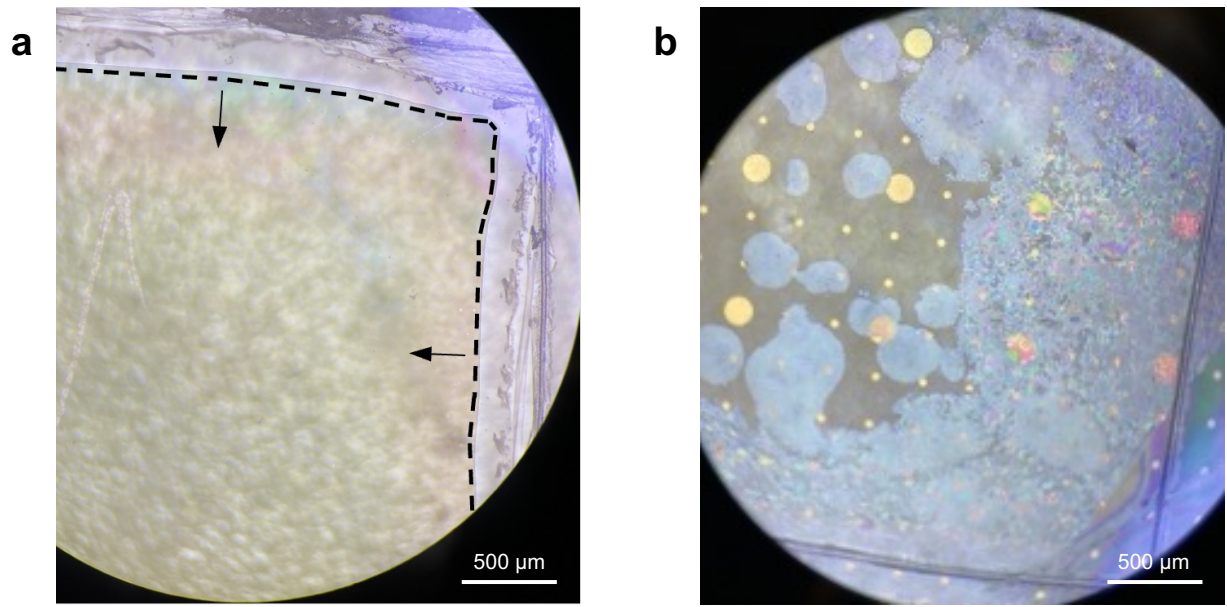

**Figure S5.** Optical microscope images of the thin film heterostructure coated with the polymer support and immersed in water. a) Uniformly advancing fronts of sacrificial layer etching observed in one of the $Nd_{0.80}Sr_{0.20}NiO_2$ heterostructure samples in the main study. The etching starts from the exposed regions near the sample edges (see Experimental Section). b) Non-uniform etching fronts and incomplete etching observed in the samples with suboptimal heterostructure templates.

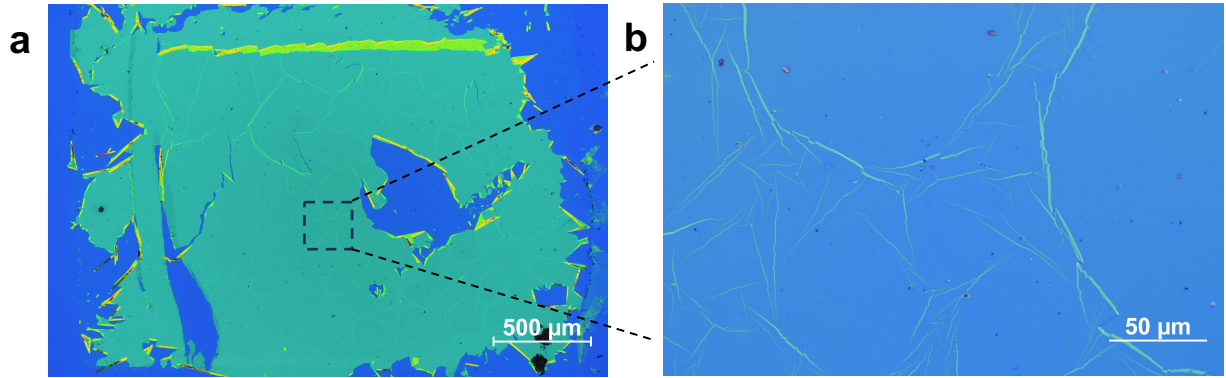

**Figure S6**. Optical microscope images of a STO / Nd$_{0.85}$Sr$_{0.15}$NiO$_2$ / STO membrane released from the NGO substrate with a lattice-mismatch of ~1.6%.

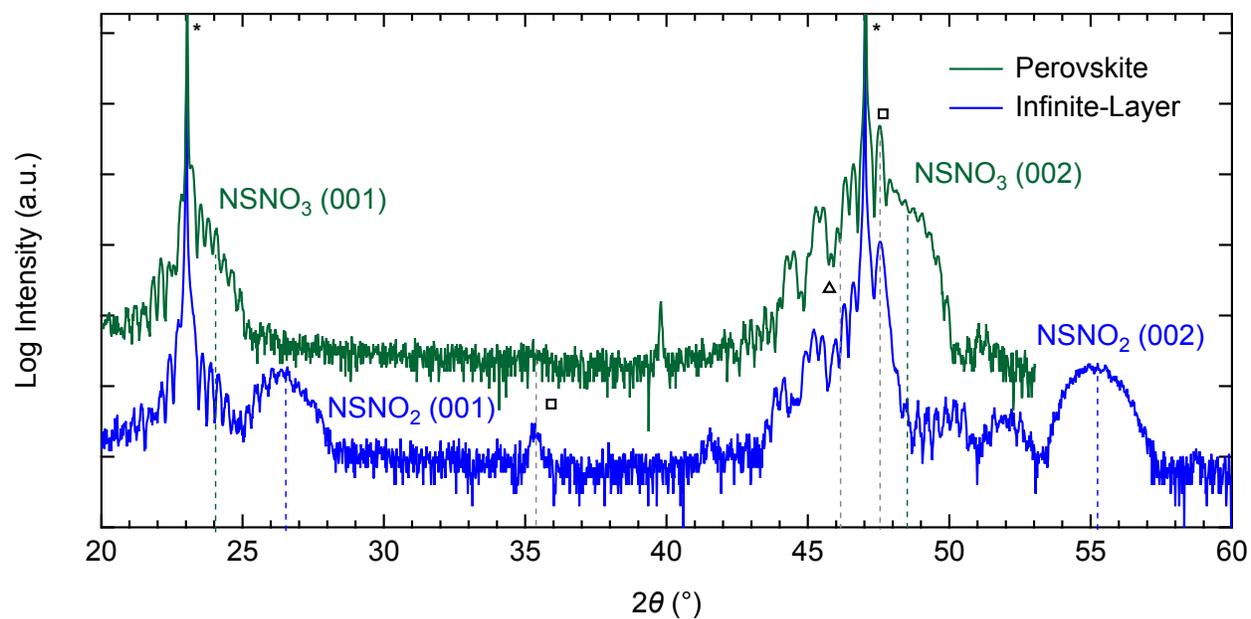

**Figure S7**. XRD $\theta$-$2\theta$ symmetric scan of 8 u.c. STO / 15 u.c. Nd$_{0.85}$Sr$_{0.15}$NiO$_3$ / 8 u.c. STO / 15 u.c. Sr$_{0.5}$Ca$_{2.5}$Al$_2$O$_6$ grown on the NGO substrate (green) and the reduced heterostructure (blue). The nickelate XRD peak positions are marked by dashed lines under the label of NSNO$_3$ (perovskite) and NSNO$_2$ (infinite-layer). *, □, and △ denote NGO, Sr$_{0.5}$Ca$_{2.5}$Al$_2$O$_6$, and STO, respectively.